\documentclass[twocolumn,showpacs,preprintnumbers,amsmath,amssymb]{revtex4}
\usepackage{graphicx}% Include figure files
\usepackage{dcolumn}% Align table columns on decimal point
\usepackage{bm}% bold math
\usepackage{latexsym}
\usepackage{amsfonts}
\usepackage{amssymb}
\usepackage{amsmath}

\begin{document}

\title{
Hawking Radiation from Squashed Kaluza-Klein Black Holes \\
--- A Window to Extra Dimensions ---
}

\author{Hideki Ishihara}
\email{ishihara@sci.osaka-cu.ac.jp}
\author{Jiro Soda}
\email{jiro@tap.scphys.kyoto-u.ac.jp}
\affiliation{
 ${}^{*}$Department of Mathematics and Physics,  Graduate School of Science,
 Osaka City University, Osaka 558-8585, Japan \\
 ${}^{\dagger}$Department of Physics,  Kyoto University, Kyoto 606-8501, Japan
}%

\date{\today}% It is always \today, today,

%===============================================================%
%************************* ABSTRACT ****************************%
%===============================================================%
\begin{abstract}
We explore the obsevability of extra dimensions through 5-dimensional
squashed Kaluza-Klein black holes residing
 in the Kaluza-Klein spacetime.
With the expectation that
 the Hawking radiation reflects the 5-dimensional nature of the squashed horizon,
we study the Hawking radiation of a scalar field in the squashed black hole background.
 As a result, we show that the luminosity of Hawking radiation tells us the size of
 the extra dimension, namely, 
  the squashed Kaluza-Klein black holes open a window to extra dimensions.
\end{abstract}

\maketitle

{\it Introduction} ---
 It is believed that the superstring theory is a promising candidate for the unified theory
 of everything. As the superstring theory can be consistently formulated only in 10-dimensions,
 the existence of extra dimensions should be regarded as the prediction of the theory.
 Therefore, it is of interest to see if these extra dimensions are observable or not.
  Here, we will seek the possibility to detect extra dimensions through black holes.
 
 As is well known,  
black holes are not black in the semiclassical theory.
Rather, they emit the Hawking radiation~\cite{Hawking:1974sw}. 
For a 4-dimensional Schwarzschild black hole, 
 the luminosity of Hawking radiation can be deduced as
$
   L = 1/1920 \pi r_{+}^2  ,
$
where we have calculated only the s-wave contribution which is dominant at low energy. 
In this conventional case, the luminosity of the Hawking radiation
is determined by the horizon size $r_{+}$.

In the higher dimensional Kaluza-Klein spacetime, it is natural to regard
black holes as the direct product of the conventional 4-dimensional 
black hole spacetime  and the internal
compact space. In the 5-dimensional case, it can be described 
by the black string. 
In this case, it is difficult to observe the extra dimension
as long as the size of the compact space is sufficiently small. 
This is because, at low energy,
the effects of the extra dimension is merely absorbed into the definition
of $r_+$. In order to see the extra dimension, we need to see the effect
where Kaluza-Klein modes play a role.

However, squashed Kaluza-Klein black holes
 recently investigated in ~\cite{Ishihara:2005dp}
 (see also \cite{Dobiasch:1981vh,Gibbons:1985ac}) have changed the situation.
The squashed Kaluza-Klein black holes  look like the 5-dimensional
black holes~\cite{Myers:1986un} in the vicinity of the horizon.
Intriguingly, the spacetime far from the black holes is locally the direct product
of the 4-dimensional Minkowski spacetime and the circle.  
In this sense, the 5-dimensional squashed Kaluza-Klein black holes
reside in Kaluza-Klein spacetime.
Since the Hawking radiation is considered as the phenomena associated
with the event horizon, it is reasonable to expect that we can observe
the extra dimension by looking at the Hawking radiation.
The purpose of this paper is to show this is indeed the case.

{\it Kaluza-Klein Black Holes with Squashed Horizons}---
Let us review the Kaluza-Klein black holes  with squashed horizons~\cite{Ishihara:2005dp}.
We consider the 5-dimensional gravity coupled with a $U(1)$ gauge field.
The action is given by 
\begin{eqnarray}
S = \frac{1}{16\pi G} \int d^5 x \sqrt{-g} 
         \left[ R - F^{\mu\nu} F_{\mu\nu} \right] ,
\end{eqnarray}
where $G$ is the 5-dimensional Newton constant, $g_{\mu\nu}$ is the metric,
and $F_{\mu\nu} = \partial_\mu A_\nu - \partial_\nu A_\mu $ is the
field strength of the 5-dimensional
$U(1)$ gauge field $A_\mu$.
We have  black hole solutions with the metric given by
\begin{eqnarray}
 ds^2 &=& -f(r) dt^2 + \frac{k^2 (r)}{f(r)} dr^2 \nonumber\\
 && \qquad\quad
 + \frac{r^2}{4} k(r) d\Omega^2    + \frac{r^2}{4} \left(d\psi + \cos \theta d\phi \right)^2 , 
\end{eqnarray}
where $d\Omega^2 = d\theta^2 + \sin^2 \theta d\phi^2$ is the metric of the unit sphere and 
\begin{eqnarray}
  f(r) = \frac{( r^2 - r_{+}^2 )( r^2 - r_{-}^2 )}{r^4}  , 
  k(r) = \frac{(r_{\infty}^2 - r_{+}^2 )(r_{\infty}^2 - r_{-}^2 )}
                  {(r_{\infty}^2-r^2 )^2} . 
\end{eqnarray}
Here, the coordinate ranges are 
$0 \leq \theta <2\pi ,  0 \leq \phi < 2\pi , 0 \leq \psi < 4\pi $. 
The gauge potential is given by
\begin{eqnarray}
     A = \pm \frac{\sqrt{3}}{2} \left( \frac{r_{+} r_{-}}{r^2} - \frac{r_{+} r_{-}}{r_\infty^2} \right)dt .
\end{eqnarray}

It is easy to see
the coordinate singularity at $r_{+}$ corresponds to the outer horizon of the black hole.
The inner horizon $r_{-}$ is analogous to that of Reissner-Nordstr\"om black holes.
The position $r_\infty$ corresponds to the spatial infinity. 
It should be noted that the shape of the horizon is a deformed sphere determined by
the parameter $k(r_{+} )$. 
The solution reveals 5-dimensional nature in the vicinity of the horizon $r_{\pm}<r \ll r_\infty$.

On the other hand, the geometry becomes the 4-dimensions with a small circle
at infinity. To see this, we use the coordinate transformation
$
   \rho = \rho_0 r^2/(r_\infty^2 - r^2 ) .
$
It yields the relation
$
  r_\infty^2  
  = 4 \left( \rho_{+} + \rho_0 \right) \left( \rho_{-} + \rho_0 \right)  ,
$
where $\rho_{\pm} = \rho_0 r_\pm^2 /(r_\infty^2 - r_\pm^2 )$.
Under this coordinate transformation, the metric reduces to
\begin{eqnarray}
 ds^2 &=& - F(\rho) d\tau^2 + \frac{K^2 (\rho)}{F(\rho)} d\rho^2 
 + \rho^2 K^2 (\rho) d\Omega^2  \nonumber\\
&& + \frac{r_\infty^2}{4 K^2 (\rho)} \left(d\psi + \cos \theta d\phi \right)^2 ,
\label{metric}
\end{eqnarray}
where
\begin{eqnarray}
F=  \left( 1- \frac{\rho_{+}}{\rho}\right)\left( 1-\frac{\rho_{-}}{\rho} \right) , \quad
K^2 =   1+ \frac{\rho_0}{\rho}  .
\end{eqnarray}
We have defined the proper time $\tau = 2\rho_0 t/r_\infty $ for the observer at infinity.
We note that even when $\rho_0$ is negative, the metric (\ref{metric}) describes 
a black hole geometry if the parameters satisfy $-\rho_0=|\rho_0|< \rho_\pm$\cite{inner}. 
In the region $\rho  \gg \rho_0 $, the geometry locally looks like the black string.
In the case $\rho_{+} < \rho_0 $, 
 the black holes look like 5-dimensional in the region $\rho < \rho_0$.  
The twist of the geometry is essential for our consideration.

Given the metric, we can calculate various physical quantities.
The surface gravity is given by
\begin{eqnarray}
   \kappa_{+} = \frac{\rho_{+} -\rho_{-}}{2\rho_{+}^2} \sqrt{\frac{\rho_{+}}{\rho_{+} + \rho_0}} .
   \label{temp}
\end{eqnarray}
Using the timelike Killing vector $\xi = \partial /\partial \tau $ normalized at
the spatial infinity, we can define the mass of the black hole as
\begin{eqnarray}
   M =  - \frac{3}{32\pi G} \int_\infty dS_{\mu\nu} \nabla^\mu \xi^\nu
     =  \frac{3\pi r_\infty}{4 G} \left( \rho_{+} + \rho_{-} \right) ,
\end{eqnarray}
where the integral is taken over the three dimensional sphere 
at the spatial infinity.
The charge of the black hole is also calculated as
\begin{eqnarray}
  Q = \frac{1}{8\pi G} \int_\infty  dS_{\mu\nu} F^{\mu\nu}
    = \frac{\sqrt{3} \pi}{G} r_\infty \sqrt{\rho_{+} \rho_{-} } .
\end{eqnarray}
We should notice that the Newton constant for the remote observer
should be
$
    G_4 = G / 2\pi r_\infty  .
$
Therefore, we cannot see the extra dimension from
the observation of the mass $M$ and the charge $Q$,
because the size of the extra dimension is absorbed into the definition of $G_4$.
We need another observable to probe the extra dimension.

{\it Mode Functions}---
Let us consider the field equation of
the massless scalar field $\Phi$ in the spacetime (\ref{metric}), 
$
\Box \Phi = 0 .
$
Fortunately, this turns out to be a separable equation. 
Taking a separable ansatz
$
  \Phi = e^{-i\omega \tau} R(\rho) e^{-i\lambda \psi} e^{i m \phi} S(\theta) ,
$
we obtain the equation
\begin{eqnarray}
  && \frac{F}{\rho^2 K^2}\frac{\partial}{\partial \rho}\left[ \rho^2 F \frac{\partial R}{\partial \rho}\right]
   + \left[ \omega^2 - \frac{4\lambda^2 K^2 F}{r_\infty^2} \right] R  \nonumber\\
   &&   \qquad\qquad = \frac{\left\{\ell (\ell+1) -\lambda^2 \right\} F}{\rho^2 K^2} R
\end{eqnarray}
for the radial part. The angular part becomes
\begin{eqnarray}
   \Delta S  -   \frac{\lambda^2 \cos^2 \theta}{ \sin^2 \theta} S
   -  \frac{2\lambda m \cos \theta}{ \sin^2 \theta} S 
    = -\left\{\ell (\ell+1) -\lambda^2 \right\} S  ,
 \label{spin}
\end{eqnarray}
where $\Delta$ is the Laplacian for the unit sphere.
The solutions for Eq.(\ref{spin}) are called the spin-weighted spherical functions.
The spectrum is determined from the regularity.
We also have conditions $\ell \geq \lambda$ and $m = -\ell, -\ell +1, \cdots, \ell$. 
The periodicity requires $2\lambda$, $2m$, and $\lambda \pm m$ are integers.

To understand the qualitative behavior of the radial mode functions,
it is convenient to use new variables
$
  d\rho_{*} = K d\rho /F , \ R (\rho) = Y (\rho)/\sqrt{\rho^2 K} .
$
The radial equations can be transformed to the Schr$\ddot{\rm o}$dinger form
\begin{eqnarray}
   - \frac{d^2 Y }{d\rho_{*}^2} + V Y = \omega^2 Y  ,
\end{eqnarray}
where
\begin{eqnarray}
  V &=& \frac{4\lambda^2}{r_\infty^2}K^2 F 
		+ \frac{(\ell (\ell +1)-\lambda^2)}{\rho^2 K^2 }F + \frac{FF'}{\rho K^2}
+ \frac{3}{4}\frac{F^2}{\rho^2 K^2} 
                                   \nonumber\\
    &&+ \frac{1}{2}\frac{F^2 K''}{K^3}
     -\frac{3}{4}\frac{F^2 K^{\prime 2}}{K^4} + \frac{1}{2} \frac{FF' K'}{K^3}.
\end{eqnarray}
\begin{figure}[h]% fig.1
%\begin{center}
\includegraphics[height=5cm, width=6cm]{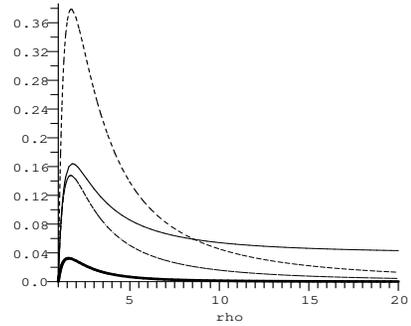}
%\end{center}
\caption{The effective potentials $V(\rho)$ for the radial modes are depicted. 
We set $\rho_{-} =0.2 , \ \rho_{+} =1$ and $\rho_0 =2$. 
The thick line represents $\ell =\lambda=0$.
The dash-dot and dash lines correspond to $\ell=1 , \lambda =0$ and $\ell =2 , \lambda =0$, respectively.
The thin line represents $\ell=1 , \lambda = 0.5$. }
\label{fig:1}
\end{figure}
From this formula, we can read off the effective potential (see Fig.1).
For higher eigenvalues $\lambda, \ell$, the height of the potential becomes  higher.
At low energy, hence, only the modes $\lambda = \ell =0$ are relevant.

{\it Hawking Radiation with Greybody Factors}---
In the semi-classical picture, the black hole can radiate particles. In fact,
the black hole can be regarded as a blackbody with the Hawking temperature $T_{BH} = \kappa_{+} /2\pi $.
The greybody factors modify the spectrum of emitted particles from
that of a perfect blackbody. 
The luminosity of the Hawking radiation with greybody factors
is then given by the formula 
\begin{eqnarray}
  L  
  =  \int_0^\infty \frac{d \omega}{2\pi} |{\cal A}(\omega )|^2
      \frac{\omega}{\exp \left( \omega / T_{\rm BH}\right)-1}  .
   \label{hawking}   
\end{eqnarray}
The boundary condition to calculate the greybody factor should be
\begin{equation}
  R(\rho)  \sim \left\{
\begin{array}{ll}
   \tilde{t} (\omega )  e^{i\omega \rho_{*}} , & \quad  \rho_{*} \rightarrow \infty   \\
   e^{i\omega \rho_{*}}  +  \tilde{r} (\omega)  e^{-i\omega \rho_{*}} , & \quad \rho_{*} \rightarrow -\infty 
\end{array}  
  \right. , 
\end{equation}
where $\tilde{t} (\omega)$ and $\tilde{r} (\omega)$ are the transmission and reflection coefficients, respectively.
However, the boundary condition for the conventional calculation is different.
The absorption probability $|{\cal A}|^2$ is computed from solutions to the wave equation
which have unit incoming flux from infinity and no outgoing flux from the past horizon:
\begin{equation}
  R(\rho)  \sim \left\{
\begin{array}{ll}
   e^{-i\omega \rho_{*}} + {\cal R} (\omega )  e^{i\omega \rho_{*}} , & \quad  \rho_{*} \rightarrow \infty   \\
   {\cal T} (\omega)  e^{-i\omega \rho_{*}} , & \quad \rho_{*} \rightarrow -\infty 
\end{array}  
  \right.  ,
\end{equation}
where ${\cal T}$ and ${\cal R}$ are, respectively, the transmission and 
reflection coefficients for this boundary condition. 
The well known relation $\tilde{t}(\omega) = {\cal T}(\omega)$ guarantees the equivalence
of both boundary conditions.
Here, we just follow this conventional procedure.
The absorption probability is the difference of the incoming and out going flux at infinity.
Thus, the absorption probability $|{\cal A}|^2$ can be calculated from the reflection
coefficient
$
  {\cal R}(\omega)  
$
as
$
  |{\cal A}|^2 =|\tilde{t}(\omega)|^2=|{\cal T} (\omega)|^2 = 1 - |{\cal R}(\omega)|^2 .
$

The wave equation we should solve is
\begin{eqnarray}
  \frac{F}{\rho^2} \frac{d}{d\rho} \left[ \rho^2 F \frac{dR}{d\rho} \right]
  +  \omega^2 \left( 1+ \frac{\rho_0}{\rho} \right) R = 0 .
  \label{basic}
\end{eqnarray}
Here, we put  $\lambda = \ell =0$ since we will consider the low energy regime.
We can not solve the above equation analytically, then we have to resort to
 an asymptotic expansion technique.
 First, we separate the region for the coordinate $\rho$
 into three parts: the near horizon $\rho_{+} \lesssim \rho $,
 the intermediate region $\rho \sim \rho_I$, and the far region $\rho \gg \rho_\pm , \rho_0$.
 Here, $\rho_I $ is some reference point where $\omega \rho_I \ll 1$ is satisfied.
 The result does not depend on the specific choice of $\rho_I$ as we will see soon. 
 We can solve the equation in each region where certain small terms are neglected.
 By matching these solutions in the overlapping regions,
 we obtain an approximate solution for the wave equation.
 Hence, we can determine the absorption probability. 

In the near horizon $\rho_{+} \lesssim \rho $, we have
\begin{eqnarray}
           \rho^2 F \frac{d}{d\rho} \left[ \rho^2 F \frac{dR}{d\rho} \right]
  +  \omega^2 \rho_{+}^4 \left( 1+ \frac{\rho_0}{\rho_{+}} \right) R = 0 .
\end{eqnarray}
Here, we have approximately put $\rho \sim \rho_{+}$ in the second term.
This can be solved by using the transformation
$
   dy / d\rho =  1/ \rho^2 F .
$
Notice that $y \sim (1/\rho_+)\log \left( \rho- \rho_{+} \right) 
\rightarrow - \infty$ at the horizon.
As we are considering the ingoing waves from the infinity,
we do not have waves coming out from the horizon, namely,
we only have ingoing waves at the horizon. 
Taking into account this ingoing boundary condition, we obtain the solution  as
\begin{eqnarray}
  R_{NH} =  a_{-} \exp\left(-i  \omega \rho_{+}^2 
		\sqrt{ 1+ \frac{\rho_0}{\rho_+}} \ y\right).
  \label{NH2}
\end{eqnarray}
In the regime $\rho > \rho_{+}$, the solution (\ref{NH2}) becomes
\begin{eqnarray}
  R_{NH} =  a_{-} \left[ 1-i  \omega \rho_{+} \sqrt{ 1+ \frac{\rho_0}{\rho_+}}
                              \log \left( \rho -\rho_{+}\right) \right].
   \label{NH}                           
\end{eqnarray}

In the intermediate zone $\rho\sim \rho_I$, Eq.(\ref{basic}) reduces to
\begin{eqnarray}
  \frac{1}{\rho^2} \frac{d}{d\rho} \left[ \rho^2  F \frac{dR}{d\rho} \right] = 0 .
\end{eqnarray}
Here, we have assumed the $\omega \rho_I \ll 1$. 
The solution is given by
\begin{eqnarray}
   R_I =  b_{+} + b_{-} \int \frac{d\rho}{\rho^2 F} .
\end{eqnarray}
In the inner region $\rho \sim \rho_+$, we obtain
\begin{eqnarray}
   R_I =  b_{+} + \frac{b_{-}}{\rho_{+}} \log \left( \rho -\rho_{+} \right) ,
   \label{In}
\end{eqnarray}
where we approximately put $\rho^2 F = \rho_+ \rho F$.
This can be matched with the near horizon solution (\ref{NH}).
In the outer region $\rho >\rho_I > \rho_+$, we can set $F\sim 1$.
 Hence, we have
\begin{eqnarray}
   R_I =  b_{+} - b_{-} \frac{1}{\rho} .
   \label{I}
\end{eqnarray}

In the far zone $\rho \gg \rho_\pm , \rho_0$, Eq.(\ref{basic}) becomes 
\begin{eqnarray}
  \frac{1}{\rho^2} \frac{d}{d\rho} \left[ \rho^2  \frac{dR}{d\rho} \right]
  +  \omega^2  R = 0 .
\end{eqnarray}
This can be solved analytically as
\begin{eqnarray}
 R_{F} = \frac{c_{+}}{\sqrt{\rho}} J_{1/2} \left( \omega \rho \right) 
 + \frac{c_{-}}{\sqrt{\rho}} N_{1/2} \left(  \omega \rho\right) .
\end{eqnarray}
In the region $ \omega \rho \ll 1$, we get
\begin{eqnarray}
 R_{F} \sim   \frac{2c_{+}}{\sqrt{\pi}} \left( \frac{ \omega}{2}  \right)^{1/2}
 - \frac{c_{-}}{\sqrt{\pi} \rho} \left( \frac{ \omega}{2} \right)^{-1/2}. 
 \label{F}
\end{eqnarray}
This can be matched with the intermediate solution (\ref{I}).
There exists the matching region $\rho_{I} < \rho < \omega^{-1}$ 
for low energy modes $\omega \rho_{I} \ll 1$.

From solutions (\ref{NH}) and (\ref{In}), we can read off the matching conditions
\begin{eqnarray}
a_{-} = b_{+}  , \quad
-i  \omega \rho_{+} \sqrt{ 1+ \frac{\rho_0}{\rho_+}} a_{-} = \frac{b_{-}}{\rho_{+}}. 
\end{eqnarray}
and, from solutions (\ref{I}) and (\ref{F}), we obtain
\begin{eqnarray}
 b_{+} =  2 c_{+} \left( \frac{ \omega}{2}  \right)^{1/2} , \quad
 - b_{-} =  - c_{-} \left( \frac{ \omega}{2} \right)^{-1/2}  .
\end{eqnarray}
In the asymptotic region $\omega \rho \gg 1$, the solution in the far region can be expanded as
\begin{eqnarray}
 R_F &=& \frac{-i c_{+} - c_{-}}{ 2 \rho} \frac{1}{\sqrt{2\pi  \omega }}
          \exp \left(i  \omega \rho \right) \nonumber\\
      &&   + \frac{ic_{+}- c_{-}}{2 \rho} \frac{1}{\sqrt{2\pi  \omega }}
  \exp \left(-i  \omega \rho\right) 
\end{eqnarray}
Therefore, the reflection coefficient is given by
$
   {\cal R} =  (ic_{-}-c_{+} )/( i c_{-}+c_{+})  . 
$
Thus, we obtain the absorption probability
\begin{eqnarray}
  |{\cal A}|^2 = 4 \left(\omega \rho_{+} \right)^2
                    \sqrt{ 1+ \frac{\rho_0}{\rho_{+}}}  .
    \label{result}                
\end{eqnarray}
This result can be understood from the behavior of the effective potential (Fig.2).
The larger $\rho_0$ yields the lower peak of the potential. Hence,
more radiations can be transmitted to the infinity.

\begin{figure}[h]% fig.2
%\begin{center}
\includegraphics[height=5cm, width=6cm]{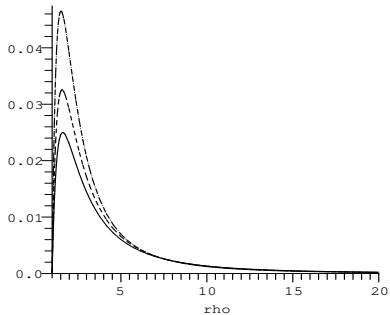}
%\end{center}
\caption{The $\rho_0$ dependence of the effective potential $V(\rho)$ is plotted.
Here, we set  $\rho_{-} =0.2 , \ \rho_{+} =1$ and $\lambda=\ell =0$.  
The dash-dot, dash, and line correspond to $\rho_0 = 1, 2, 3$ respectively.     }
\label{fig:2}
\end{figure}

{\it Discussion}---
 We have considered a scalar field in Kaluza-Klein black hole with the squashed horizon
and obtained  the greybody factor (\ref{result}).
Now, combining Eqs. (\ref{temp}), (\ref{hawking}), and (\ref{result}),
 we have the luminosity of the Hawking radiation
\begin{eqnarray}
   L = \frac{1}{1920 \pi \rho_{+}^2 } \left(1-\frac{\rho_{-}}{\rho_{+}}\right)^4 
   \left(1+ \frac{\rho_0}{\rho_{+}} \right)^{-3/2}  .
   \label{main}
\end{eqnarray}
In the the limit, $\rho_0 \rightarrow 0$, we obtain the metric 
(\ref{metric}) with $K^2 =1$ which locally has the geometry of the black string.
In this limit, Eq.(\ref{main}) reduces to the standard formula for
the 4-dimensional charged black holes.
Therefore, without squashed horizons, we never notice
the extra dimension even if we can detect Hawking radiations. 
For black holes with $\rho_0 \gg \rho_+ $, the luminosity of Hawking radiation
is significantly suppressed compared with that of the conventional 4-dimensional black holes.
For the negative $\rho_0$, it is highly enhanced. This $\rho_0$ dependence
allows us to see the extra dimension. 

Now we come to our conclusion. 
Three observable quantities $M,Q$ and $L$ determine $\rho_{\pm}$ and $\rho_0$.
Therefore, we can calculate the size of the extra dimension $ r_{\infty}$.
Thus, in principle, we can observe the extra dimension
through the Hawking radiation. Namely, the squashed black holes could be
a window to extra dimensions. 

Although we have discussed a 5-dimensional example, squashed higher dimensional
black holes residing in higher dimensional Kaluza-Klein spacetime
might exist. Attempts to find explicit solutions are important.  
It is of interest  to extend our analysis to more general 
situations~\cite{Wang:2006nw,Yazadjiev:2006iv,Brihaye:2006ws,
Ishihara:2006ig,Ishihara:2006pb,Ishihara:2006iv}.
We also need to study the stability of the squashed Kaluza-Klein black holes. 
There may exists the Gregory-Laflamme type instability~\cite{Gregory:1993vy}.
Moreover, we can consider the squashed black holes in the braneworld context.
Although inhomogeneities in the perpendicular direction to the brane and the radion
make the situation complicated~\cite{Tamaki:2003bq}, the possibility of twists 
must provide us rich physical phenomena.
\begin{acknowledgements}
H.I. is supported by the Grant-in-Aid for  Scientific
Research Fund of the Ministry of Education, Science and Culture of Japan 
  No.13135208. 
J.S. is supported by 
the Japan-U.K. Research Cooperative Program, the Japan-France Research
Cooperative Program and Grant-in-Aid for  Scientific
Research Fund of the Ministry of Education, Science and Culture of Japan 
  No.18540262 and No.17340075.  
\end{acknowledgements}

\end{document}